\documentclass[aps,prl,reprint,showpacs,superscriptaddress,groupedaddress]{revtex4-1}

\usepackage{graphicx}
\usepackage{amsmath}
\usepackage{amssymb}
\usepackage{dcolumn}
\usepackage{dsfont}
\usepackage{latexsym}
\usepackage{rotating}
\usepackage{color}
\usepackage{latexsym}
\usepackage{bbm}
\usepackage{subfigure}
\usepackage{float}
\usepackage{epsfig}
\usepackage{psfrag}
\usepackage{natbib}
\usepackage{bm}
\usepackage{amsthm}
\usepackage{eucal}
\usepackage{mathrsfs}

\usepackage{color}

\usepackage{hyperref}
\hypersetup{
colorlinks=true,final=true,
        linkcolor=blue,
        citecolor=blue,
        filecolor=blue,
        urlcolor=blue,
}
\begin{document}

\title{Oxygen-vacancy   clustering    and   pseudogap   behaviour   at
  LaAlO$_3$/SrTiO$_3$ interface}

\author{N. Mohanta} \email{nmohanta@phy.iitkgp.ernet.in}
\author{A. Taraphder} \email{arghya@phy.iitkgp.ernet.in}
\affiliation{Department of Physics and Centre for Theoretical Studies,
  Indian Institute of Technology Kharagpur, W.B. 721302, India}

\begin{abstract}
  The 2DEG at the LaAlO$_3$/SrTiO$_3$  interface promises to add a new
  dimension to emerging electronic devices due to its high tunability.
  Defects in  the form  of Oxygen vacancies  in titanate  surfaces and
  interfaces, on the  other hand, play a key role  in the emergence of
  novel phases.  Based  on an effective model, we  study the influence
  of  Oxygen  vacancies  on  superconductivity and  ferromagnetism  at
  LaAlO$_3$/SrTiO$_3$  interface.    Using  the  Bogoliubov-de  Gennes
  formulation in  conjunction with  Monte-Carlo simulation, we  find a
  clustering  of the Oxygen  vacancies at  the interface  that favours
  formation  of coexisting  ferromagnetic puddles  spatially separated
  from  superconductivity. We also  find a  carrier freeze-out  at low
  temperatures,  observed experimentally  in wide-variety  of samples.
  Sufficiently large amount of Oxygen vacancy leads to pseudo-gap like
  behaviour in the superconducting state.
 
\end{abstract}

% Vacancies, in crystals, 61.72.jd
% Superconducting films, 74.78.-w 
% Magnetic properties of thin films, surfaces, and interfaces spin-orbit 
% effects, 75.70.Tj
% Phase separation and segregation in thin films, 64.75.St
\maketitle

\section{Introduction}
The discovery of a two-dimensional electron gas (2DEG) at the (001)-interface of
two  oxide insulators  SrTiO$_3$  and LaAlO$_3$~\cite{Ohtomo2004}  is hailed  as
unique,  due  to  the  gamut  of   novel  properties  that  it  has  thrown  up:
low-temperature             superconductivity             (below             200
mK)~\cite{Reyren31082007,0953-8984-21-16-164213},         coexistence         of
superconductivity   and   ferromagnetism~\cite{Li2011,  PhysRevLett.107.056802},
electric                      field-induced                      metal-insulator
transition~\cite{Cen2008,Caviglia2008,lu:172103}   and  superconductor-insulator
transition~\cite{PhysRevLett.103.226802}.  The  2DEG at the  interface is formed
as a result  of an intrinsic electronic transfer process  occurring to nullify a
charge-discontinuity                            at                           the
interface~\cite{Nakagawa2006,PhysRevB.80.075110,Warusawithana2013}.  The 2DEG is
lying in  a material which is  more complex than simple  semiconductors and thus
the ground state phase-daigram consists of several competing phases which can be
turned ON  or OFF by changing some  external parameters such as  gate voltage or
magnetic   field.    The   fantastic   tunability   of  the   2DEG   makes   the
LaAlO$_3$/SrTiO$_3$  interface  an  important  material  for  future  electronic
devices~\cite{Mannhart26032010,mannhart_apl2012,ref1}.      However    intrinsic
defects such as the Oxygen vacancies, created during the deposition process, can
influence several properties and  thus deserve serious attention while designing
such   devices.    The   oxygen    vacancies   contribute   electrons   to   the
interface~\cite{PhysRevB.75.121404,PhysRevLett.98.196802, PhysRevLett.98.216803}
that    are    responsible     for    metallic    conduction    for    amorphous
samples~\cite{PhysRevX.3.021010,Schlom2011}.   Annealing in  presence  of Oxygen
destroys  these  vacancies,  degrading  the  mobility  of  carriers  while  both
crystalline  and amorphous  samples,  which  are not  annealed,  show a  carrier
freeze-out  below  100~K~\cite {PhysRevX.3.021010,PhysRevLett.107.146802}.   The
critical  LaAlO$_3$ thickness  (4 unit  cells for  crystalline samples)  for the
appearance   of   metallicity   decreases    with   the   increase   in   Oxygen
vacancies~\cite{PhysRevX.3.021010}.
 
Apart  from  the  modulation  of  electronic  properties,  the  Oxygen
vacancies  have significant  influence on  the magnetic  properties as
well.   Density functional theory  suggests that  magnetism is  not an
intrinsic property  of the  interface electrons, it  is caused  by the
spin-splitting  of  the populated  electronic  states  induced by  the
Oxygen     vacancies     in     the     SrTiO$_3$     or     LaAlO$_3$
layer~\cite{PhysRevB.85.020407}.    Oxygen-annealing   results  in   a
quenching  of magnetic  moment,  observed in  X-ray magnetic  circular
dichroism (XMCD) measurement~\cite{2013arXiv1305.2226S}, an indication
that  the  electrons  donated  by  Oxygen  vacancies  are  essentially
localized at the interface.   According to Pavlenko, et al. \textit{et
  al.}~\cite{2013arXiv1308.5319P},   the  Oxygen   vacancies   at  the
interface lead to an interplay of ferromagnetic order in the itinerant
$t_{2g}$   bands   and    complex   magnetic   oscillations   in   the
orbitally-reconstructed $e_g$ bands.

However, it  is important to study  the effect of  Oxygen vacancies on
the  superconducting state  as well,  since a  competing ferromagnetic
order is  present with non-zero vacancy  concentration.

Our analysis,
based on  a BdG  formalism followed by  Monte-Carlo simulation  in the
superconducting state reveals a  clustering of Oxygen vacancies in the
two-dimensional (2D) plane at very low-temperatures. Superconductivity
is  affected by  the Oxygen  vacancies in  a non-trivial  way.  In the
clustered   regions,  superconductivity   is  destroyed   locally  and
ferromagnetic puddles are formed.  As we argue here, the career-freeze
out effect observed in the  samples might be related to the clustering
of the  vacancies within  the interface. The  local density  of states
(LDOS) shows  a strong dependence on vacancy  concentration along with
the emergence of a pseudo-gap like behaviour.

\section{Model and method}
The interface electrons occupy the  $t_{2g}$ bands of Ti atoms and are
responsible for ferromagnetism and superconductivity at the interface.
The $d_{xy}$ band  has a bandwidth twice that  of $d_{xz}$ or $d_{yz}$
and is located relatively lower in energy at the $\Gamma$ point. Hence
electrons occupying this band are  expected to be localized due to the
Coulomb repulsion at the interface. These localized electrons interact
with the  conduction electrons via a  ferromagnetic exchange, modelled
by     a     Zeeman     field     parallel    to     the     interface
plane~\cite{Lee2013,2013arXiv1309.1861M}.      In     fact,     recent
spectroscopic studies  revealed direct evidence  for ferromagnetism of
$d_{xy}$  character  in  the plane~\cite{Lee2013}.   Superconductivity
originates from usual  phonon-mediated electron-electron pairing as in
doped          SrTiO$_3$          substrates~\cite{PhysRevLett.14.305,
  jourdan2003superconductivity}.  The highly broken inversion symmetry
at  the  interface plane  leads  to  a  large Rashba  spin-orbit  (SO)
interaction which  modifies the electronic bands  and pairing symmetry
of the superconducting state~\cite{2013arXiv1309.1861M}.  The in-plane
Zeeman   field  brings   about   an  assymetry   in  the   two-sheeted
Fermi-surface  (created   by  the   SO  interaction)  and   favours  a
finite-momentum pairing of the electrons.

Motivated by this phenomenology  and electronic structure, we consider
the following  Hamiltonian for a  Rashba SO coupled  superconductor in
presence  of a  uniform in-plane  Zeeman field  and  Oxygen vacancies~
\cite{Lee2013,2013arXiv1309.1861M}:
\begin{equation}
\begin{split}
{\cal H}&=-t\sum_{<ij>,\sigma}(c_{i\sigma}^\dagger c_{j\sigma}+h.c.)-
\sum_{i,\sigma}(\mu-V\delta_i) c_{i\sigma}^\dagger c_{i\sigma}\\
&-H_x\sum_{i,\sigma,\sigma^{\prime}}(\sigma_x)_{\sigma \sigma^{\prime}}c_{i\sigma}^\dagger c_{i\sigma^{\prime}}+ \sum_{i}\Delta(r_i)(c_{i\uparrow}^{\dagger}c_{i\downarrow}^{\dagger}+h.c.)\\
&-i\frac{\alpha}{2}\sum_{<ij>,\sigma,\sigma^{\prime}}c_{i\sigma}^{\dagger}(\boldsymbol{\sigma}_{\sigma \sigma^{\prime}} \times \boldsymbol{d}_{ij})_z c_{j\sigma^{\prime}}
\end{split}
\label{model}
\end{equation}
\noindent where  $t$ is  the hopping  amplitude (set to  1 as  unit of
energy)  on a  2D square  lattice, $H_x$,  chosen along  $x$ direction
without loss of generality, is  the in-plane Zeeman field. $\alpha$ is
the  Rashba SO  coupling strength,  $\boldsymbol{d}_{ij}$ is  a vector
connecting  sites $i$  and  $j$, and  $\boldsymbol{\sigma}$ being  the
Pauli matrices.  The parameter  $\delta_i$ represents the  location of
Oxygen  vacancies,  is $1$  for  sites  with  Oxygen vacancy  and  $0$
otherwise,  describing a  defect state  as a  local shift  $V$  in the
chemical  potential  $\mu$. In  the  following  analysis,  we work  at
half-filling,  as superconductivity  has maximum  stability  there (we
checked  at other fillings).  It is  also assumed  that the  effect of
Oxygen  vacancy  is the  same  at all  vacancy  sites  and take  $V=2$
throughout.   We  checked  that  the  underlying  physics,  under  the
approximations,  is qualitatively  similar for  a range  of  values of
$V$.  The  local  superconducting  pairing  amplitude  is  defined  as
$\Delta(r_i)=-<c_{i\uparrow}c_{i\downarrow}>$.

The Hamiltonian (\ref{model}) is diagonalized using a spin-generalized
Bogoliubov-Valatin transformation
$\hat{c}_{i\sigma}(r_i)=\sum_{i,\sigma^{\prime}}u_{n\sigma\sigma^{\prime}}(r_i)
\hat{\gamma}_{n\sigma^{\prime}}+v_{n\sigma\sigma^{\prime}}^*(r_i)
\hat{\gamma}^{\dagger}_{n\sigma^{\prime}}$   which  gives the  local
pairing gap in terms of Bogoliubov amplitudes $u_{n\sigma}(r_i)$ and 
$v_{n\sigma}(r_i)$ as
\begin{equation}
\begin{split}
\Delta(r_i)&=-\sum_{n}[u_{n\uparrow}(r_i)v^*_{n\downarrow}(r_i)(1-f(E_n))\\
&+u_{n\downarrow}(r_i)v^*_{n\uparrow}(r_i)f(E_n)]
\end{split}
\label{del}
\end{equation}
\noindent  where $f(x)=1/(1+e^{x/{k_BT}})$  is the  Fermi  function at
temperature  $T$.  The in-plane  magnetization density  and occupation
number are obtained via
\begin{equation}
\begin{split}
m(r_i)&=\frac{1}{2}<c_{i\uparrow}^{\dagger}c_{i\downarrow}+c_{i\downarrow}^{\dagger}c_{i\uparrow}>\\
&=\sum_{n,\sigma}u_{n\sigma}^{*}u_{n\sigma^{\prime}}f(E_n)+v_{n\sigma}u_{n\sigma^{\prime}}^{*}(1-f(E_n))
\end{split}
\end{equation}
\vspace{-1em}
\begin{equation}
\begin{split}
n(r_i)&=\sum_{\sigma}<c_{i\sigma}^{\dagger}c_{i\sigma}>\\
&=\sum_{n,\sigma}(|u_{n\sigma}|^2f(E_n)+|v_{n\sigma}|^2(1-f(E_n)))
\end{split}
\end{equation}

\noindent   The  quasi-particle   amplitudes   $u_{n\sigma}(r_i)$  and
$v_{n\sigma}(r_i)$ are determined by solving the BdG equations
\begin{equation}
\begin{split}
{\cal H}\phi_n(r_i)=\epsilon_n\phi_n(r_i)
\end{split}
\label{bdg eq.}
\end{equation} 
where
$\phi_n=[u_{n\uparrow}(r_i),u_{n\downarrow}(r_i),v_{n\uparrow}(r_i),
v_{n\downarrow}(r_i)]$.   \\

The local order parameters  are solved self-consistently on a finitely
large square lattice with periodic boundary conditions and finally the
average  values are  calculated. The  Oxygen vacancy  configuration is
annealed   by  a   Monte-Carlo  simulation,   starting  from   a  high
temperature, in which the free-energy is updated using the eigenvalues
of  the BdG  Hamiltonian~\cite{}.   As temperature  goes down,  larger
number  of Monte-Carlo steps  were performed,  each step  allowing one
Oxygen vacancy to move to one of its nearest neighbor non-vacant sites
following the standard Metropolis algorithm (being charged impurities,
long range  transfer of vacancies  is highly unlikely, even  at higher
temperatures).   Each  MC  step  then  involves solving  the  the  BdG
equations  followed by  a random  change in  vacancy  configuration as
above.   At low  temperatures, the  superconducting  and ferromagnetic
properties are calculated.

\section{Results and discussions}
\begin{figure}[!ht]
\epsfig{file=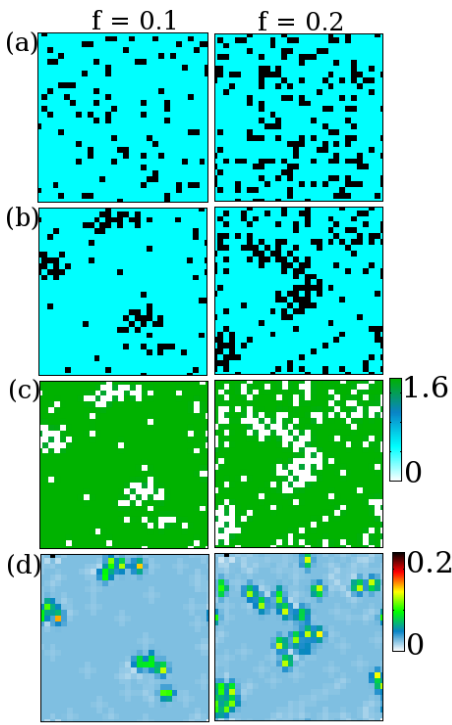, width=0.7\linewidth}
\caption{(Color  online)  (a)   Initial  and  (b)  final  vacancy
  configurations with  vacancy concentration  $f = 0.1$  (left column)
  and $f=0.2$ (right  column).  Panel (c) and (d)  are the profiles of
  local     pairing     amplitude     and     onsite     magnetization
  respectively. Parameters used:  $\mu=0$, $\alpha=0.8$ and $H_x=0.5$.}
\label{order}\vspace{-0.5em}
\end{figure}

The Monte-Carlo annealing shows a tendency towards bunching of the 
Oxygen vacancies, as shown in FIG.~\ref{order}, panel (a),(b), which
turns out to be crucial in establishing a ferromagnetic order.  
One Oxygen vacancy adds two additional electrons to the interface which are 
then pinned near the vacant sites. However, a considerable concentration of 
vacancies  are required  for the  emergence of overall ferromagnetism~
\cite{2013arXiv1309.1861M,2013arXiv1308.5319P}. Such defect-induced ferromagnetism 
is also observed in bulk SrTiO$_3$ substrate 
attesting to the connection between the level of impurity and  observed
ferromagnetism~\cite{PhysRevX.2.021014,crandles:053908}. In the
clustered regions, the pairing gap collapses due to large fluctuations
in local  electron density and the localized  moments, originated from
the Oxygen vacancies, order to form ferromagnetic puddles, as shown in
FIG.~\ref{order}(c)  and (d).  The clustering  of vacancies  implies a
spatial  phase  separation  of  superconductivity  and  ferromagentism
allowing them  to coexist at the interface~\cite{Wang2011}  and can be
detected  by   positron  annihilation   spectroscopy  or  NMR   as  in
Ref.~\onlinecite{1999SSCom.112..245E,PhysRevB.76.172106}.

The  Oxygen-vacancies disrupt the lattice and effectively act as
localized centres with enhanced scattering of electrons, which
is stronger at higher temperatures. At high temperatures, predictably, 
there is no evidence of the ordering or bunching of the vacancies.
\begin{figure}[!ht]
\epsfig{file=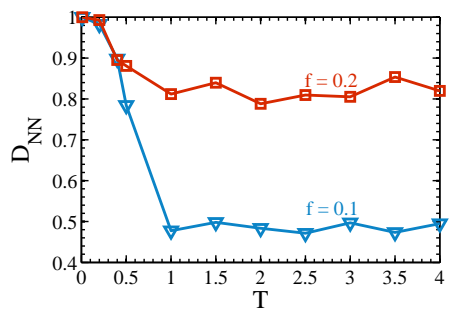, width=0.7\linewidth}
\caption{(Color online) Temperature  variaton of the nearest-neighbour
  vacancy-vacancy  correlation  function  (normalized  to  unity)  for
  vacancy  concentrations   $f=0.1$  and  $f=0.2$.    Parameter  used:
  $\mu=0$, $\alpha=0.8$ and $H_x=0.5$.}
\label{vac_corr}\vspace{-0.5em}
\end{figure} 
\noindent At low temperatures, the  vacancies tend to form clusters as
depicted  in  FIG.~\ref{vac_corr},   where  we  plot  the  temperature
variation  of nearest-neighbour  vacancy-vacancy  correlation function
$D_{NN}=\frac{1}{N}  \sum_{<ij>}V_i \cdot  V_j$,  where $V_{i,j}$  are
impurity  potentials at  sites $i,j$  respectively. This  behaviour is
more prominent with sufficiently  large number of vacancies.  
Mannhart \textit{et  al.}~\cite{PhysRevB.86.235418} found,  by  second harmonic 
generation   experiment,  that   the  impurity-signal   maximizes  for
LaAlO$_3$ film of  thickness 1 u.c. and proposed a possible explanation 
via uncompensated polar distorion. The above  results show
that  a clustering  of the  Oxygen  vacancies may be possible within  the
two-dimensional TiO$_2$ layer even without a polar distortion. 
With increasing vacancy concentration $f$, the mean pairing gap reduces 
while the average  magnetic moment increases, as shown in FIG.~\ref{vac_var}.
\begin{figure}[!ht]
\epsfig{file=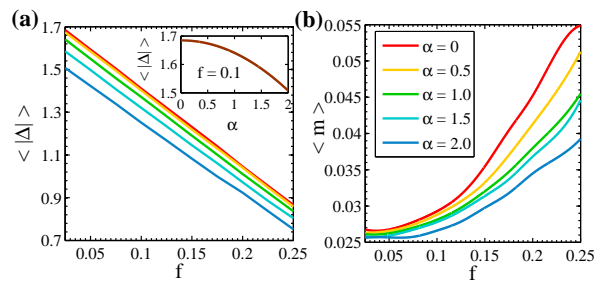, width=0.9\linewidth}
\caption{(Color online) The variation of (a) local superconducting gap
  parameter   and  (b)  onsite   magnetization  with   Oxygen  vacancy
  concentration for different SO strengths. Inset in (a) shows
  the  variation   of  mean  pairing  gap   with  SO  coupling
  strength. Parameter used: $\mu=0$ and $H_x=0.5$.}
\label{vac_var}\vspace{-0.5em}
\end{figure} 
At  a  finite  vacancy  concentration,  both  superconducting  pairing
(FIG.~\ref{vac_var}   (a)) and ferromagnetic order
(FIG.~\ref{vac_var}(b))  are  quite  sensitive  to  Rashba  SO
interaction,  since  the spin-precession  due  to  Rashba coupling  is
strongly affected by scattering from the defects created by the Oxygen
vacancies.  The   temperature  dependence  of   superconductivity  and
ferromagnetism   with   changing    Oxygen   vacancy   is   shown   in
FIG.~\ref{temp_var}. Below a  superconducting transition at $T_c=0.8$,
both superconductivity and ferromagnetic order coexist.
\begin{figure}[!ht]
\epsfig{file=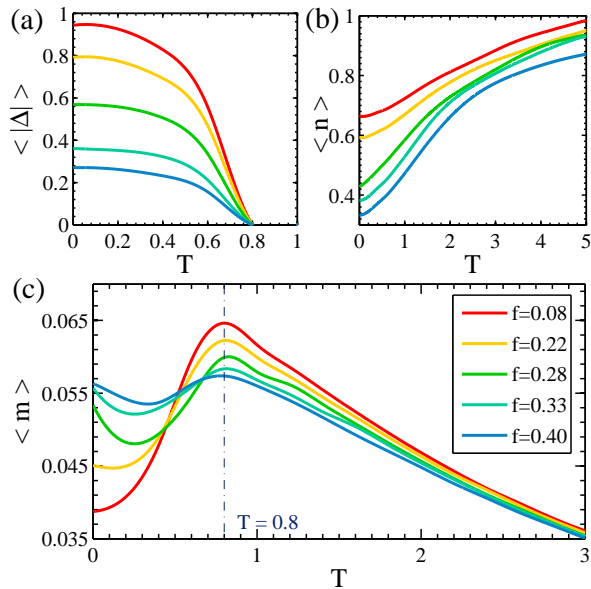, width=0.9\linewidth}
\caption{(Color online) The temperature variations of (a) mean pairing
  gap, (b) average occupation number and (c) average magnetization for
  a  range of vacancy  concentrations.  Parameter  used: $\alpha=0.8$
  and $H_x=0.5$.}
\label{temp_var}\vspace{-0.5em}
\end{figure} 
The average magnetization shows a peak at $T_c$ which is determined by
the interplay of superconductivity  and ferromagnetism and reflects an
electronic  phase  separation.  Below  $T_c$,  there  is  inhomogenous
mixture of  superconductivity and ferromagnetism  in mutually excluded
regions  in the  plane and  magnetism appears  in the  form  of robust
ferromagnetic  puddles  in the regions of the  clustering  of  Oxygen
vacancies, where superconductivity has been degraded. Thus  the 
average moment increases  with increasing vacancy
concentration. On  the other hand, beyond $T_c$,  the in-plane Zeeman
field gives  rise to a ferromagnetic order  which pervades
the  two-dimensional plane.  The long  tail  in FIG.~\ref{temp_var}(c)
implies  that the  ferromagnetism is  extended upto  temperatures well
beyond     $T_c$, which  has also been reported
experimentally~\cite{PhysRevLett.107.056802}.   Both  crystalline  and
amorphous interfaces, as well as Oxygen-deficient SrTiO$_3$ substrates 
exhibit a carrier freeze-out below $\sim$100 K~\cite{PhysRevX.3.021010, 
PhysRevB.74.035112,Brinkman2007} following the empirical relation  
$n  \propto   e^{-\epsilon/{k_BT}}$,  up  to  very  low
temperatures,  where $\epsilon$  is the  activation energy  for Oxygen
vacancy formation  (about 4.2  meV for Oxygen-deficient  SrTiO$_3$ and
0.5  meV  for  Oxygen-annealed  LaAlO$_3$/SrTiO$_3$  interface).  This
carrier-freezing  effect is very  important and  sometimes leads  to a
metal-insulator transition~\cite{PhysRevLett.107.146802}. We observe a
similar behavior for the carrier concentration in our analysis as the 
temperature is lowered, shown in FIG.~\ref{temp_var}(b).  Clearly, the 
$T=0$ value of carrier concentration is strongly defect-limited as well.

\begin{figure}[!ht]
\epsfig{file=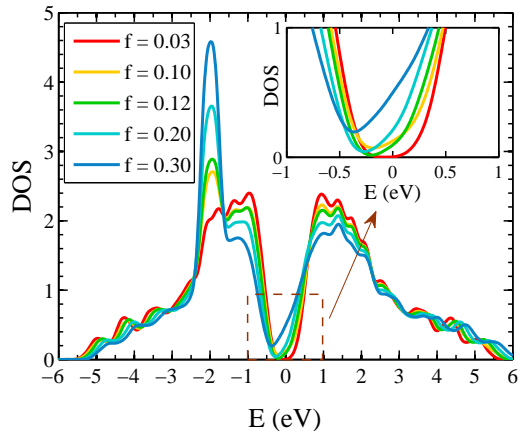, width=0.8\linewidth}
\caption{(Color  online) The  density of  states for  different Oxygen
vacancy concentrations,  showing the appearance of  pseudogap states at
higher vacancy concentrations. Parameter used: $\mu=0$, $\alpha=0.8$
and $H_x=0.5$.}
\label{dos}\vspace{-0.5em}
\end{figure}

Superconducting properties depend sensitively on the stoichiometry and
subsequent ordering  of Oxygen vacancies  as reported by  some earlier
experimental studies~\cite{Jorgensen1988578,Williams1989331}.  Oxygen
vacancies destroy  the superconducting order locally (FIG.~\ref{order}(c)).  
In FIG.~\ref{dos}, we show the density of states
of the system for a  range of vacancy concentrations. 

As   Oxygen   vacancy  concentration   is   increased  gradually   the
superconducting energy gap closes  and a pseudo-gap like feature shows
up.  Recent  tunnelling spectra measurement~\cite{Richter2013} revealed  
a pseudo-gap  behaviour of the interface  superconductivity similar to
the  high-T$_c$  cuprates.  Since  Oxygen  vacancy  is an inalienable 
part of SrTiO$_3$ substrates,  the pseudo-gap  is expected  to be
present  in any  SrTiO$_3$-based interface  supercondutor  with broken
inversion symmetry.  With increasing vacancy concentration, there is a
pile up  of the states at around  -2 eV (Fig. 5)reflecting the  pinning 
at the vacancy potential; Oxygen vacancy  affects all vacancy sites uniformly
in our model (\ref{model}). There is a gradual shift of the states to
lower energies with increasing Oxygen vacancy and in realistic situations the 
peak around -2 eV will get broadened due to the spread in vacancy potential.

\section{Conclusion}
Oxygen vacancy at  the interface is a key  ingredient of the LaAlO$_3$/SrTiO$_3$
interface and plays  a major role in the  emergence of coexisting ferromagnetism
and  superconductivity.   The  Monte-Carlo  method  coupled  to  BdG  mean-field
formalism  captures the  essential features  of the  inhomogeneous  system.  The
Rashba  SO interaction, originated  from the  highly broken  inversion symmetry,
creates helical  bands in  which the electron  spins are aligned  to essentially
render  the  superconductivity a  chiral  $p_x  \pm  ip_y$-wave one.   Following
Michaeli \textit{et al.}~\cite{PhysRevLett.108.117003}, we assume that there are
localized  electrons from  the $d_{xy}$  band  acting as  localized moments  and
interact with the conduction electrons  via a ferromagnetic exchange (modeled as
a uniform Zeeman field parallel to the interface plane). The conventional wisdom
is that  electrons cannot be both ferromagnetically  ordered and superconducting
simultaneously. Here, the defects in the form of Oxygen vacancies play a crucial
role in stabilizing a robust ferromagnetism and hence the coexistence of the two
competing  orders.   Even though,  there  can  certainly  be effects  of  Oxygen
vacancies  at the  layers  just above  or  below the  interface  plane which  is
TiO$_2$-terminated, our calculations show that  a clustering of the vacancies in
the  2D   plane  at   very  low  temperatures   is  possible   with  significant
implications. It  helps establish long-range ferromagnetic order  in the regions
where superconductivity is  locally degraded. It also sheds  light on the career
freeze-out effect  observed experimentally as the temperature  is reduced.  With
increasing vacancy  concentration, pseudogap states appear in  the local density
of states.  The  analysis, presented here, is useful  for a detail understanding
of the  role of  Oxygen vacancies in  the properties  of field-effect
devices fabricated using LaAlO$_3$/SrTiO$_3$ heterostructures.

\section*{Acknowledgements}
This  work was  supported by  Ministry of  Human  Resource Development
(MHRD)  and  Council of  Scientific  and  Industrial Research  (CSIR),
India.

%\bibliography{interface}

\end{document}